\documentstyle[12pt,epsfig]{article}
\begin{document}
\thispagestyle{empty}
\setcounter{page}0

~\vfill
\begin{center}
{\Large\bf	On the CVC problem in $\tau$-decay} \vfill

{\large	M. V. Chizhov}
 \vspace{1cm}

{\em Center of Space Research and Technologies, Faculty of Physics,
University of Sofia, 1164 Sofia, Bulgaria}

\end{center}  \vfill

\begin{abstract}

The preliminary results of PIBETA experiment strongly suggest the presence 
of non $V\!-\!A$ anomalous interactions in the radiative pion decay.  
We assume that they arise as a result of the exchange of new intermediate 
chiral spin-1 bosons which interact anomalously with matter.
Their contribution into the $\tau$-decay leads to violation of the CVC 
hypothesis at the same level as detected experimentally. 

\end{abstract}

\vfill

\newpage


The $\tau$ lepton weak decay into the $\rho$ meson $\tau^-\to\nu\rho^-$
is a pure vector process. On the other hand the $\rho$ meson can be
produced in $e^+ e^-$ annihilation through a virtual photon
$e^+ e^-\to\gamma^*\to\rho^0$. 
According to the Standard Model (SM) the $W$-boson and the photon
originate from the same multiplet of $SU_W(2)$ group and they have 
the same form of vector coupling with matter. Therefore,
these two processes of the $\rho$ meson production should be related
by the CVC hypothesis~\cite{CVC}.

In order to estimate hadronic contribution 
$a_\mu^{\rm had}$ into the muon anomalous 
magnetic moment a comprehensive analysis of $e^+ e^-$ and $\tau$
data has been fulfilled~\cite{DEHZ}. Since the main contribution in 
$a_\mu^{\rm had}$ comes from the $\rho$-resonance region 
the $\pi^+\pi^-$ channel has been thoroughly investigated.
After taking into account all known isospin breaking corrections
it was found that the $\tau$-data lead to a bigger value of
$a_\mu^{\rm had}$. The reason of that is discrepancy (up to 10\%), 
mainly above the $\rho$-resonance region, in spectral
functions $v_0(s)$ and $v_-(s)$ extracted from $e^+ e^-$~\cite{old,CDMnew} 
and $\tau$~\cite{tau} data. Despite of larger errors
this discrepancy is even more pronounced in comparison with 
the older $e^+ e^-$ experiments~\cite{old} data
which show systematically lower values of the spectral function $v_0(s)$
than $v_-(s)$.

To find solution of this problem
the authors~\cite{iso} have assumed that the problem 
is mainly due to additional isospin breaking effects and not to 
experimental ones. The solution suggested in this letter is more radical
and is based on present experimental data. The anomalous 
results~\cite{ISTRA,PIBETA} on the radiative pion decay 
$\pi\to e\nu\gamma$ (RPD), which
also have  problem with CVC hypothesis, definitely lead to prediction
of the discussed effect in the \mbox{$\tau$-decay}. In other words the
destructive interference in RPD and excess of the $\rho$ meson production
in the \mbox{$\tau$-decay} are due to the presence of new {\it centi-weak}
tensor interactions.

The idea of possible presence of such interactions is very simple and 
natural~\cite{review}.
They could be induced by an exchange of new {\it chiral} spin-1 particles,
which are different from the gauge ones. The introduction of an additional
new 
type of spin-1 particles follows from the fact, that in the relativistic
theory there are {\it two inequivalent} representations 
of the Lorentz group with spin 1. All known spin-1 particles, the photon, 
$Z$, $W$, and the gluons, are described by the gauge fields $A_\mu$, which are
transformed under the vector representation (1/2,1/2). Obviously, 
the antisymmetric tensor fields $T_{\mu\nu}$ transforming under the other 
inequivalent representation \mbox{$(1,0)+(0,1)$} can describe unknown yet 
spin-1 particles. They just lead to effective tensor interactions. 

Such type of  chiral spin-1 bosons can be identified among the hadron 
resonances~\cite{NJL}. Owing to the unique quantum numbers $1^{+-}$
($J^{PC}$) $b_1(1235)$ meson just represents a pure state 
of the chiral axial-vector 
particle interacting anomalously with the tensor quark current 
$\partial_\nu(\bar{q}\sigma_{\mu\nu}\gamma^5 q)$.
It was shown also that the $\rho(770)$ and $\rho'(1450)$ mesons are 
two orthogonal states consisting of mixture of pure 
gauge-like and chiral vector particles with near maximal mixing. 
This  leads to a number of predictions, namely  the mass formula 
\begin{equation}
3m^2_{b_1}=2m^2_{\rho'}-m_{\rho'}m_{\rho}+2m^2_{\rho},
\label{mass}
\end{equation}
which is fairly satisfied also for the spin-1 isosinglets $I=0$
$\omega(782)$,
$\omega'(1420)$, $h_1(1170)$ and $\phi(1020)$, $\phi'(1680)$, $h_1(1380)$; 
 the ratio 
\begin{equation}
\frac{f^T_\rho}{f_\rho}\simeq\frac{1}{\sqrt{2}}
\label{ratio}
\end{equation}
of the $\rho$ meson couplings to vector and tensor currents defined as
\begin{eqnarray}
\langle\rho^-(q)\vert\bar{d}\gamma_\mu u\vert 0\rangle&=&
m_\rho f_\rho \varepsilon^*_\mu,\nonumber\\
\langle\rho^-(q)\vert\bar{d}\sigma_{\mu\nu} u\vert 0\rangle&=&
i f^T_\rho \left(q_\mu\varepsilon^*_\nu-q_\nu\varepsilon^*_\mu\right),
\label{coupling constants}
\end{eqnarray}
which is in good accordance with the QCD sum rules~\cite{QCD} 
and the lattice calculations~\cite{lattice}.  

Now it is natural to assume that the same phenomenon may take place on more
fundamental level. One of the possibilities  to incorporate
the chiral spin-1 particles into SM, has been presented in~\cite{MPL}.
Then the effective Lagrangian of quark-lepton tensor interactions reads
\begin{equation}
{\cal L}_T =  -f_T\frac{G}{\sqrt{2}}~ 
\bar{u}\sigma_{\lambda\alpha}d~
\frac{4q_\alpha q_\beta}{q^2}~
\bar{e}\sigma_{\lambda\beta}(1-\gamma^5)\nu_e +{\rm h.c.},
\label{int}
\end{equation}
where $q_\mu$ is the momentum transfer between quark and lepton currents.
The dimensionless constant $f_T$ determines the relative strength of the new
tensor interactions with respect to the ordinary weak interactions and
it is predicted to be positive.
Using the result of ISTRA experiment on RPD~\cite{ISTRA} and applying
the QCD sum rules method with PCAC technique~\cite{SUSY}
one estimates the tensor coupling constant as $f_T \sim 0.02$~\cite{mu}.

To explain the result of high statistics and high precision
PIBETA experiment~\cite{PIBETA} on RPD two times smaller value 
of the tensor coupling constant $f_T \simeq 0.01$ is necessary.

\begin{figure}[th]
\mbox{\hspace{2cm}}\epsfig{file=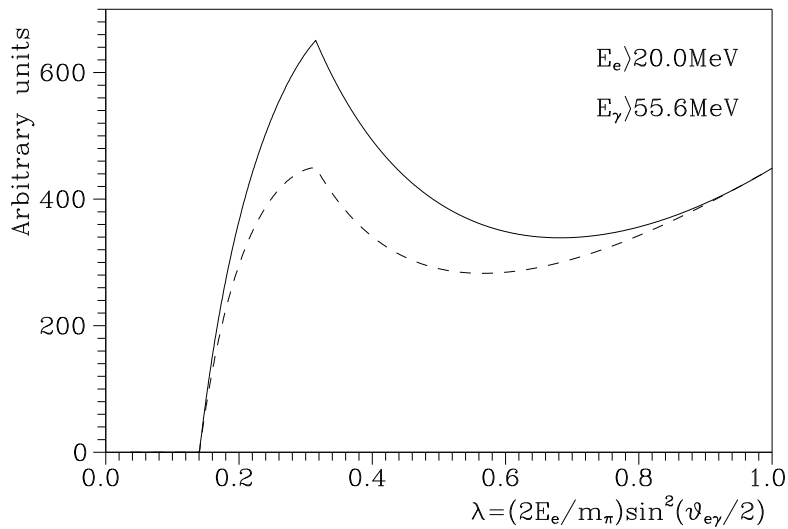,height=7cm,width=10.5cm}
\caption{Theoretically predicted RPD yield is shown by the  solid curve.
The dashed curve presents the effect of the new tensor interactions
(\ref{int})
with $f_T=0.01$.}
\label{lambda}
\end{figure}

\noindent In the Fig.~\ref{lambda} the qualitative agreement for the
deficit 
in PRD yield with \cite{PIBETA} is shown.
In other words the new tensor interactions are two orders of magnitude
weaker
than ordinary weak interactions 
and present new class of {\it centi-weak} interactions. 

Although the chiral fields are transformed as doublets under $SU_W(2)$ and
the gauge fields, corresponding to $W$-bosons, 
are transformed as triplets, they can mix after
the symmetry breaking like the $\rho-\rho'$ mixing in hadron physics.
If the chiral bosons are very heavy then it is a unique possibility for
them to manifest themselves through a mixing with $W$-bosons. This case was
considered in~\cite{zurich}. It seems that the case of pure mixing, neglecting
the interactions (\ref{int}), is disfavored by the present data. However, a
combined case may be revealed by a precise fit of PIBETA data.

To estimate an effect of the new tensor interactions on $\tau$-decay
the Lagrangian (\ref{int}) will be used. The matrix element of
$\tau$-decay into $\rho$ meson can be obtained using definitions 
(\ref{coupling constants})
\begin{equation}
{\cal M}=-\frac{G}{\sqrt{2}}m_\rho f_\rho~\varepsilon^*_\mu(q)
~\bar{\nu}_\tau(1+\gamma^5)\gamma_\mu\tau
-4if_T\frac{G}{\sqrt{2}} f^T_\rho~q_\mu\varepsilon^*_\nu(q)
~\bar{\nu}_\tau(1+\gamma^5)\sigma_{\mu\nu}\tau.
\label{M}
\end{equation}
Then the ratio of the spectral functions $v_-(s)$ and $v_0(s)$ at
the $\rho$-resonance region
\begin{equation}
\frac{v_-(s)}{v_0(s)}=1+F_T~\frac{6s}{m^2_\tau+2s}
+F^2_T~\frac{s(2m^2_\tau+s)}{m^2_\tau(m^2_\tau+2s)},
\label{result}
\end{equation}
where
\begin{equation}
F_T=4f_T~\frac{m_\tau}{m_\rho}~\frac{f^T_\rho}{f_\rho}>0,
\end{equation}
should be greater than one even in the limit of exact isospin invariance.

Since the $\rho$ and $\rho'$ mesons are the mixture of gauge-like and chiral
vector states the coupling constants $f_\rho$ and $f^T_\rho$ are not constants
anymore and they depend on $s=q^2$ in a specific way~\cite{NJL}. 
Explicit forms of these dependencies can help, probably, to specify more 
accurately resonance fitting  parameters. For a simple
evaluation of the effect of the new tensor interactions one can use
the coupling constants ratio (\ref{ratio}) at $s=m^2_\rho$. 
It leads to 5.5\%
excess in the $\rho$ production in $\tau$-decay with respect to CVC 
prediction. This value is in good agreement with the ratio
\begin{equation}
{B^{\rm exp}(\tau^-\to\nu_\tau\pi^-\pi^0)
- B^{\rm CVC}(\tau^-\to\nu_\tau\pi^-\pi^0)
\over B^{\rm CVC}(\tau^-\to\nu_\tau\pi^-\pi^0)} = 6.2\pm 1.4\% 
\end{equation}
derived from the older data~\cite{old} and does not contradict 
$3.8\pm 1.4\%$ when the new data~\cite{CDMnew} are used.

Therefore, the introduction of the new tensor interactions (\ref{int})
allows to explain both the significant deficit in RPD yield and the excess
of the $\rho$ mesons in $\tau$-decay with respect to CVC predictions.
These effects are noteworthy now  because they are statistically
significant and 
based  on several experiments and many experimental data.

\pagebreak[4]

\end{document}